\documentclass[epj]{svjour}

\newcommand{\whizard}{\texttt{WHIZARD}}
\newcommand{\GeV}{{\ensuremath\rm GeV}}
\newcommand{\TeV}{{\ensuremath\rm TeV}}
\newcommand{\chp}{\tilde{\chi}^+}
\newcommand{\chm}{\tilde{\chi}^-}

\newcommand{\ME}{\mathcal{M}}
\newcommand{\ab}{{\ensuremath\rm ab}}
\newcommand{\eqn}{equation}
\newcommand{\formcalc}{\texttt{FormCalc}}
\usepackage{graphicx}
\usepackage{fancyhdr}
\usepackage{amssymb,amsmath,array}
\def\makeheadbox{{
 }}
\def\mytitle{My title} 
\def\myauthors{My name}  
\def\mytype{My type of session}
\def\mysession{My session}

\def\mytitle{NLO Event Simulation for ILC Chargino Production} 
\def\myauthors{Tania Robens}    
\def\mytype{Contributed Talk}    
\def\mysession{Colliders - SUSY Phenomenology}

\begin{document}
\title{NLO Event Simulation for Chargino Production at the ILC}
\author{Tania Robens
\thanks{\emph{Email:} robens@physik.rwth-aachen.de}%
 }
%
%
\institute{RWTH Aachen - Institut f\"ur Theoretische Physik E -
52056 Aachen - Germany\vspace{2mm} \\
PITHA 07/12,\hspace{2mm}SFB/CPP-07-67
}
%
\date{}
\abstract{ We present an extension of the Monte Carlo Event Generator \whizard~
 which includes chargino production at the ILC at NLO. We include photons using both a fixed order and a resummation approach. While the fixed order approach suffers from negative event weights, the resummation method solves this problem and automatically includes leading higher order corrections. We present results for cross sections and event generation for both methods and evaluate the systematic errors due to soft and
  collinear approximations.  In the resummation approach, the residual
  uncertainty can be brought down to the per-mil level.\\
    This is an updated version of the results presented in \cite{Kilian:2006bg,Robens:2007hn}.
} 
\maketitle
\section{Introduction}
\label{intro}
In many GUT models, the masses of charginos tend to be near the lower
edge of the superpartner spectrum, and  they can be pair-produced at a first-phase ILC with c.m.\ energy of
$500\;\GeV$. The precise measurement of their parameters (masses,
mixings, and couplings) is a key for uncovering 
the fundamental properties of the
MSSM~\cite{Aguilar-Saavedra:2005pw}. Regarding the
experimental precision at the ILC, off-shell kinematics for the signal process, and
the reducible and irreducible backgrounds~\cite{Hagiwara:2005wg} need to be included as well as NLO corrections for chargino
production at the ILC which are in the percent regime. We here present the inclusion of the latter.
\section{Chargino production at LO and NLO}
\subsection*{Fixed order approach}
The total fixed-order NLO cross section is given by
\begin{eqnarray}
\lefteqn{\sigma_{\rm tot}(s,m_e^2) =}\\
&& \sigma_{\rm Born}(s) + 
  \sigma_{\rm v+s}(s,\Delta E_\gamma,m_e^2) + 
  \sigma_{\rm 2\rightarrow 3}(s,\Delta E_\gamma,m_e^2),\nonumber
\end{eqnarray}
where $s$ is the cm energy, $m_{e}$ the electron mass, and
$\Delta\,E_{\gamma}$ the soft photon energy cut dividing the photon
phase space.  
The 'virtual' 
contribution $\sigma_\text{v}$ is the interference of the one-loop corrections
\cite{Fritzsche:2004nf} with the Born term. The collinear and infrared
singularities are regulated by $m_e$ and the photon mass $\lambda$, respectively. 
The dependence on $\lambda$ is eliminated by
adding the soft real photon contribution $\sigma_{\rm s}
\,=\,f_{\rm soft}\,\sigma_{\rm Born}(s)$ with a universal soft factor 
$f_{\rm soft}(\frac{\Delta E_\gamma}{\lambda})$
\cite{Denner:1991kt}. We break the `hard' contribution 
$\sigma_{\rm 2\rightarrow 3}(s,\Delta E_\gamma,m_e^2)$, i.e., the
real-radiation process $e^-e^+\rightarrow\chm_i\chp_j\gamma$,
into a
collinear and a non-collinear part, separated at a photon
acollinearity angle $\Delta\theta_\gamma$ relative to the incoming
electron or positron. The collinear part is approximated by convoluting the Born cross section with a
structure function\\ $f_{\rm h}(x;\Delta\theta_\gamma,\frac{m_e^2}{s})$
\cite{Bohm:1993qx}:
\begin{eqnarray}
\lefteqn{\sigma_\text{h,c}({\textstyle \Delta E, \Delta \theta, s})=}\nonumber\\
&&\hspace{10mm}\int_{\Delta E,0}^{E_\text{max},\Delta\theta}\,dx_{i}\,d\Gamma_{2}\,f_{\rm h}(x_{i})|\ME_{b}|^{2}(x_{i},s),
\end{eqnarray} 
where $x_{i}$ denotes the momentum fraction of the respective incoming beam after photon radiation.
The non-collinear part is generated
explicitely using exact three particle final state kinematics.
 
The total fixed order cross section 
is implemented in the multi-purpose event generator
\whizard~ \cite{Moretti:2001zz,Kilian:2007gr} using 
a `user-defined' structure 
function and an effective matrix
element 
\begin{eqnarray}
\lefteqn{|\ME_\text{eff}|^2=}\\
&& (1+f_{s}(\Delta E_{\gamma},\,\lambda))\,|\ME_{\text{born}}|^{2}\,+\,2\,Re(\ME_{\text{born}}\,\ME_{\text{virt}}^{*}(\lambda))\nonumber
\end{eqnarray}
which contains the
Born part, the soft-photon factor and the Born-1 loop interference
term.  
\begin{figure}
\centering
\includegraphics[width=.32\textwidth]{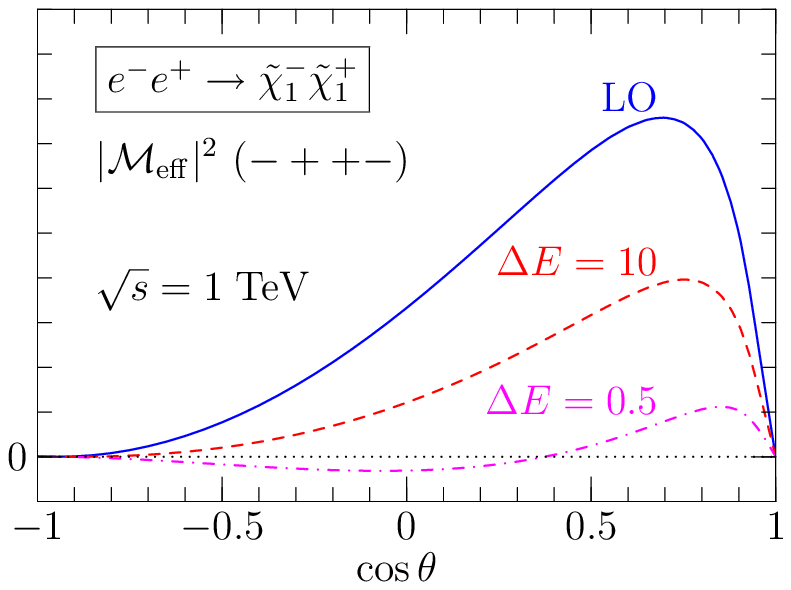} \quad
\includegraphics[width=.32\textwidth]{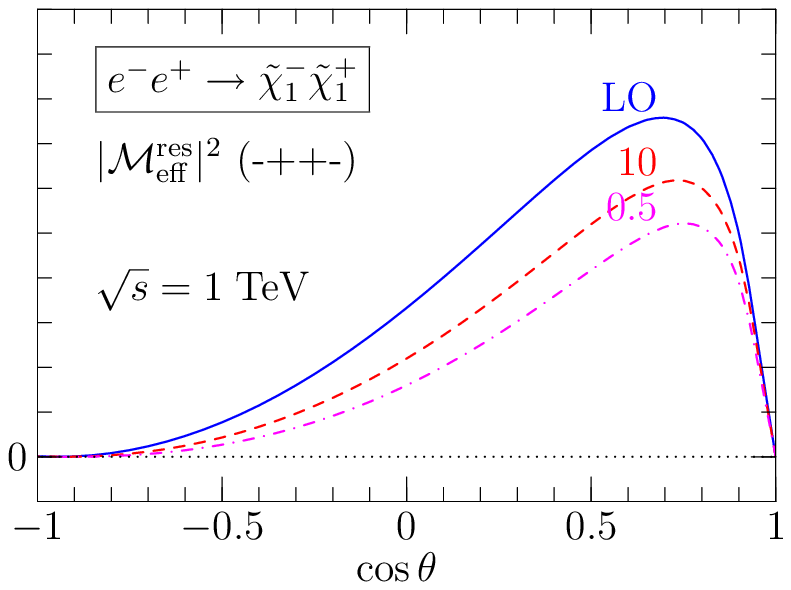}
\caption{{\small $\theta$ -dependence of effective squared matrix element 
  ($\sqrt{s}=1\;\TeV$).Upper figure: fixed order effective matrix element; lower figure: effective matrix element with
  the one-photon ISR part subtracted.  Blue/ solid line: Born term; red/ dashed: including virtual
  and soft contributions for $\Delta E_\gamma=10\;\GeV$; magenta/ dotted: same with
  $\Delta E_\gamma=0.5\;\GeV$.  
  $\Delta\theta_\gamma=1^\circ$.}}
\label{fig:Meff}
\end{figure}
In the soft-photon region this approach runs into
the problem of negative event
weights~\cite{Kleiss:1989de}: for some values of $\theta$,  
the $2\to 2$ part of the NLO-corrected
squared matrix element is positive definite by itself only if $\Delta
E_\gamma$ is sufficiently large, cf Fig.~\ref{fig:Meff}.
To still obtain unweighted event samples, an ad-hoc approach is to simply drop events with negative events before
proceeding further.
\subsection*{Resummation approach}
Negative event weights can be avoided by
resumming
higher-order initial
radiation using an exponentiated structure function 
$f_\text{ISR}$~\cite{Gribov:1972rt,Skrzypek:1990qs}. In order to avoid dou-\\ble-counting in the 
combination of the ISR-resummed LO result with the additional NLO 
contributions \cite{Fritzsche:2004nf}, we have subtracted from the effective
squared matrix element the soft and virtual photonic contributions that have already been
accounted for in $\sigma_\text{s+v}$.
This defines 
\begin{\eqn}
|\ME^{\text{res}}_\text{eff}|^2 \, = \,
|\ME_\text{eff}|^{2}- 2f_\text{soft,ISR}  \,|\ME_\text{Born}|^2 
\end{\eqn}
which is positive for even low $\Delta E_{\gamma}$ cuts for all values of $\theta$ (cf Fig. \ref{fig:Meff}). 
 Convoluting
this with the resummed ISR structure function for each incoming beam,
we obtain a modified $2\to 2$ part of the total cross section which contains all NLO contributions and in addition includes higher order soft and collinear photonic 
corrections to the Born/one-loop interference. This differs from the standard treatment in the literature (cf eg. \cite{Fritzsche:2004nf}) where higher order photon contributions are combined with the Born term only (``Born+''). 

The complete result
also contains the hard non-collinear $2\to 3$ part convoluted with
the ISR structure function:
\begin{eqnarray}
\lefteqn{\sigma_{\text{res,+}}=\int^{\Delta (E,\theta)} \,dx_{i}\,d\Gamma_{2}\,f_{\text{ISR}}^{(e^{+})}(x_{1})f_{\text{ISR}}^{(e^{-})}(x_{2})  |\ME^{\text{res}}_{\text{eff}}|^{2}}\nonumber\\
&&+\int_{\Delta (E,\theta)} dx_{i}\,d\Gamma_{3}\,f_{\text{ISR}}^{(e^{+})}(x_{1})f_{\text{ISR}}^{(e^{-})}(x_{2})|\ME^{2\rightarrow\,3}|^{2}\nonumber\\
&&
\end{eqnarray}
The resummation
approach eliminates the problem of negative weights
(cf.~Fig.~\ref{fig:Meff}) such that unweighting of generated events and
realistic simulation at NLO are now possible in all regions of
phase-space.
\section{Results}
\subsection*{Total cross section and relative corrections}
\begin{figure}
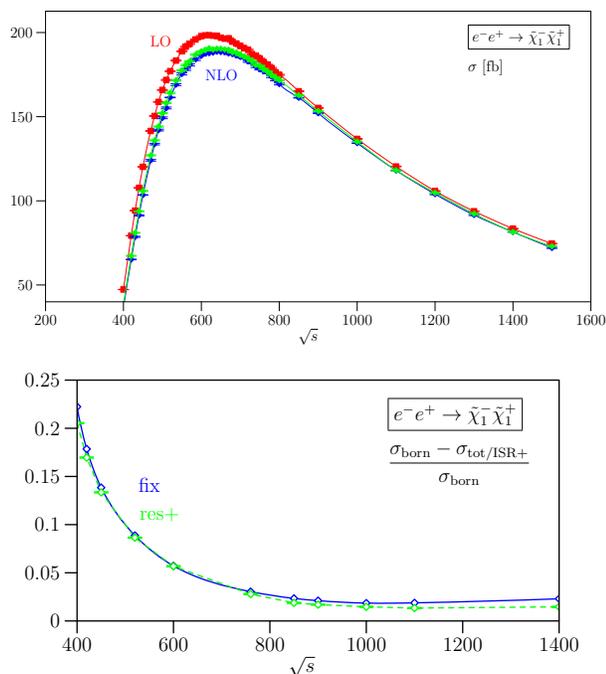

\centering
\includegraphics[width=.43\textwidth]{sigresum} 
\vspace{10mm} \quad
\includegraphics[width=.38\textwidth]{resumreldif} \vspace{5mm}
\caption{Total and relative cross section as a function of $\sqrt{s}$. Upper figure: Born (red, ``LO''), fixed order (blue, ``NLO'') and fully resummed (green, ``NLO'') total cross section, lower figure: relative  fixed order (blue, solid) and fully resummed (green, dashed) higher corrections with respect to Born result}
\label{fig:sigtot}
\end{figure}
Figure \ref{fig:sigtot} shows the c.m. energy dependence of the total LO and NLO cross section for chargino production for the mSugra point SPS1a' \cite{Aguilar-Saavedra:2005pw} and the relative corrections with respect to the Born result. The corrections are mostly in the percent regime and can reach $20\%$ in the threshold region.
\subsection*{Cutoff dependencies}
\begin{figure}
\centering
  \includegraphics[height=0.3\textwidth,width =0.5\textwidth]{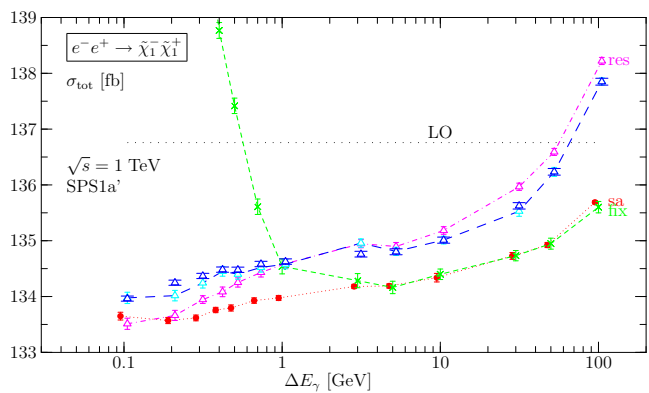}
  \caption{\label{fig:edep} {\small Total cross section dependence on $\Delta
    E_\gamma$: {\rm `sa'}
    (red, dotted) = fixed-order semianalytic result; {\rm `fix'} (green, dashed) = fixed-order
    Monte-Carlo result; {\rm `res'}
    (blue, long-dashed) = ISR-resummed Monte-Carlo result; (magenta, dash-dotted) = same but resummation applied only to
    the $2\to 2$ part. $\Delta\theta_\gamma=1^\circ$.
    LO: Born cross section.}}
\label{fig:edep}
\end{figure}
 Fig.~\ref{fig:edep} compares the $\Delta E_{\gamma}$ dependence of
 the numerical results from 
 a semianalytic fixed-order calculation with the Monte-Carlo
integration in the fixed-order and in the resummation schemes. The fixed-order Monte-Carlo result agrees with the semianalytic
result as long as the cutoff is greater
than a few $\GeV$ but departs from it for smaller cutoff values
because here, in some parts of phase 
space, $|\ME_{\rm eff}|^{2}\,<\,0$ is set to zero. The semianalytic
fixed-order result is not exactly 
cutoff-independent, but exhibits 
a slight rise of the calculated cross section with increasing cutoff due to the breakdown of the soft photon approximation. For $\Delta E_\gamma=1\;\GeV$
($10\;\GeV$) the shift is about 
$2\,\text{permil}$ ($5\,\text{permil}$) of the total cross section. The fully resummed result  shows an increase of about
$5\,\text{permil}$ of the total cross section with respect to the
fixed-order result which stays roughly constant until $\Delta
E_\gamma>10\;\GeV$.  This is due to higher-order photon radiation.

For the dependence on the collinear cutoff $\Delta\theta_\gamma$, the main higher-order effect is associated with
photon emission angles below $0.1^\circ$. For $\theta_\gamma>10^\circ$,
the collinear approximation breaks down. 
\subsection*{Event distributions}
\begin{figure*}
\centering
  \includegraphics[width=.6\textwidth]{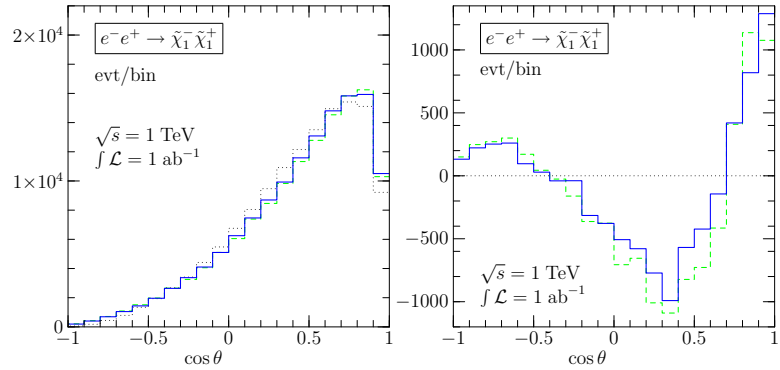}
  \caption{{\small Polar scattering angle distribution for an integrated
    luminosity of $1\;\ab^{-1}$ at $\sqrt{s}=1\;\TeV$. Left: total
    number of events per bin; right: difference w.r.t.\ the Born
    distribution.  LO (black, dotted) = Born cross section without
    ISR; fix (green, dashed) = fixed-order approach; res (blue, full)
    = resummation approach.  Cutoffs: $\Delta E_\gamma=3\;\GeV$ and
    $\Delta\theta_\gamma=1^\circ$.}}
\label{fig:histth}
\end{figure*}

 In
Fig.~\ref{fig:histth} we show the binned distribution of the chargino
production angle obtained using a sample of unweighted events. 
It demonstrates that NLO corrections to the angular distribution are statistically significant and cannot be
accounted for by  a constant K factor. 

\begin{figure*}
\centering
\includegraphics[height=0.22\textwidth, width=.36\textwidth]{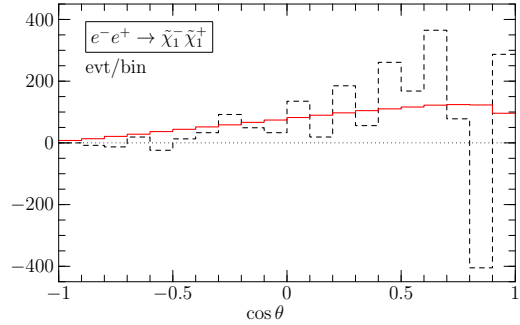} 
\vspace{5mm}
\caption{Polar scattering angle dependence of difference between events resulting from completely resummed and fixed order method: $N_{res}\,-\,N_{fix}$. Standard deviation from Born (red, solid) is shown}
\label{fig:evhigho}
\end{figure*}

Figure \ref{fig:evhigho} shows the difference between the angular event distribution resulting from the fixed order method and the resummation method. As a reference, one standard deviation from the Born event distribution is also given. It is clearly visible that, in the central-to-forward region, even higher order contributions are statistically significant.

%
\subsection*{$\sqrt{s}$ dependence of higher order contributions}
\begin{figure*}
\centering
  \includegraphics[height=0.25\textwidth, width =0.5\textwidth]{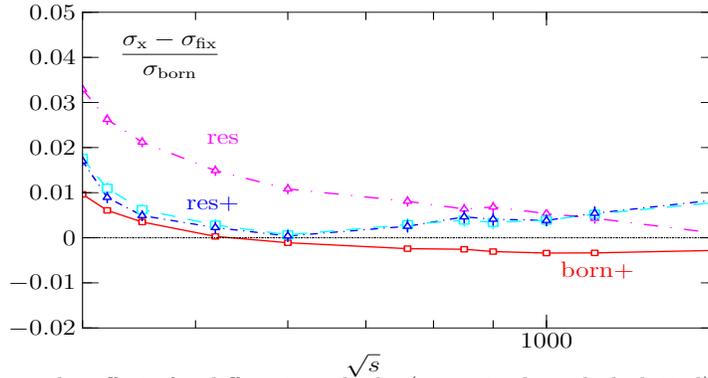}
\vspace{4mm}
  \caption{{\small  Relative higher-order effects for different methods: (magenta, long dash dotted) = $\sigma_{\text{res}}$, (blue/ cyan and dash-dotted/ dashed) =  $\sigma_{\text{res}+}$, and (red, solid) = $\sigma_{\text{Born}+}$  vs $\sigma_\text{Born}$}.    }
\label{fig:secoeff}
\end{figure*}

Figure \ref{fig:secoeff} shows the magnitude of second and higher order photonic effects in different schemes. We compare the fully resummed version (``res+'') to the standard way of adding higher order photonic corrections in the literature (``Born+''), which consists in only convoluting the Born contribution with $f_\text{ISR}$, and a not completely resummed version (``res'') where higher order photon radiation is taken into account in the $2\rightarrow 2$ kinematic regions only. Comparison of ``res'' and ``res+'' shows that the additional convolution of $\ME^{2\rightarrow\,3}$ leads to corrections in the percent regime.
Comparing ``Born+'' and ``res+'', we see that, for $\sqrt{s}\,>\,500\GeV$, the additional convolution of the interference term with $f_\text{ISR}$ is equally in the percent regime and additionally changes the sign of the higher order corrections. Therefore, all higher order contributions which are contained in ``res+'' are significant and have to be taken into account.
 For more details, cf.~\cite{Kilian:2006cj,Robens:2006np}.

\section{Conclusions}
We have implemented NLO corrections
into the event generator \whizard~for chargino pair-production at the
ILC with several approaches for the inclusion of photon radiation. 
A careful analysis of the dependence on the cuts
$\Delta\,E_{\gamma},\,\Delta\,\theta$  reveals 
uncertainties related to higher-order radiation and breakdown of the
soft or collinear approximations. To carefully choose the
resummation method and cutoffs will be critical for a truly precise
analysis of real ILC data.The version of the program resumming
photons allows to get rid of negative event weights,
accounts for all yet known \\higher-order effects, allows for 
cutoffs small enough that soft- and collinear-approximation artefacts
are negligible, and explicitly generates photons where they can be
resolved experimentally. Corrections for the decays of charginos \cite{Rolbiecki:2007cv,Fujimoto:2007bn} and
non-factorizing corrections are in the line of future work. Equally, a general interface to \formcalc~\cite{Hahn:1998yk} in principle allows for the inclusion of any lepton-collider NLO production process in \whizard. 
\section*{Acknowledgements}
This work was supported by the DFG SFB/TR9
"Computational Particle Physics" and German Helmholtz Association, Grant VH-NG-005.
%
 \bibliographystyle{unsrt}
 \bibliography{susy07_tr}
%

\end{document}